\documentclass[aps,prd,reprint,nofootinbib,superscriptaddress,floatfix,longbibliography]{revtex4-2}
\usepackage[T1]{fontenc}
\usepackage{lmodern}
\usepackage{amsmath,amssymb,amsthm,bm,booktabs,mathrsfs,graphicx}
\usepackage{xcolor}
\usepackage[colorlinks=true,linkcolor=blue,citecolor=magenta,urlcolor=blue]{hyperref}
\usepackage{orcidlink}
\newcommand{\dd}{\mathrm{d}}
\newcommand{\R}{\mathbb{R}}
\newcommand{\Ai}{\operatorname{Ai}}
\newcommand{\atanh}{\operatorname{artanh}}
\newcommand{\A}{\mathcal{A}}

\newcommand{\tcell}[1]{\begingroup\raggedright #1\par\endgroup}

\begin{document}
\title{Exact particle count from relativistic moving mirrors}
\author{Michael R. R. Good\,\orcidlink{0000-0002-0460-1941}}
\email{muon@asu.edu}
\affiliation{Physics Department \& Energetic Cosmos Laboratory, Nazarbayev University, Astana 010000, Qazaqstan}
\affiliation{Leung Center for Cosmology \& Particle Astrophysics, National Taiwan University, Taipei 10617, Taiwan}
\affiliation{Beyond Center for Fundamental Concepts in Science, Arizona State University, Tempe, AZ 85287, USA}

\author{Evgenii Ievlev\,\orcidlink{0000-0002-5935-4706}}
\email{evgeniiievlev@cuhk.edu.cn}
\affiliation{School of Science \& Engineering, Chinese University of Hong Kong, Shenzhen, 518172, China}

\author{Yedige Zholdasbay\,\orcidlink{0009-0000-1814-609X}}
\email{zholdasbay.edu@gmail.com}
\affiliation{EdVille International School, Astana 010000, Qazaqstan}

\begin{abstract}
For a real massless scalar field in two-dimensional flat spacetime, we obtain exact particle counts for families of perfectly reflecting mirrors that start and end at rest. The trajectories differ in how rapidly their speed falls away from its maximum. 
At equal maximum speed and a common time scale, changing the shape of motion can increase the number of particles while reducing their total energy. In the ultrarelativistic limit, we compute general particle count scaling and test it on our exactly solvable examples.
\end{abstract}
\maketitle

\section{Introduction}\label{sec:introduction}
It is fascinating that an accelerating boundary can create particles from the vacuum \cite{FD1976,DF1977}. This model, also called a moving mirror model, belongs to the wider study of quantum fields in time-dependent backgrounds \cite{Birrell}, with moving cavity walls providing an early formulation of boundary-induced particle creation \cite{Moore1970}. Recent developments include a geometrical approach to constructing exact solutions for dynamical cavities~\cite{Araujo2026}. A mirror boundary mixes positive and negative field frequencies, and the resulting radiation can be described by its particle spectrum and by its stress-energy flux. However, these descriptions answer different questions. Energy measures the frequency-weighted radiation; particle number counts the quanta without that weight. A large increase in energy can therefore come from more particles, more energetic particles, or both.

Black-hole radiation and the response of accelerated observers provide related settings in which the definition of particles depends on the spacetime evolution or on the observer \cite{Hawking1975,Unruh1976}. Explicitly solvable moving-mirror models make part of this relation explicit \cite{CarlitzWilley, Fernandez-Silvestre:2021ghq, Good:2018ell, Good:2020fjz, Good:2016yht}. 
In this paper, we propose several examples of exactly solvable relativistic moving mirrors and analytically compute their total particle production. We then prove general scaling properties of energy and particle counts and use our examples to test them.

This particle production is also known within the dynamical Casimir effect, which has a broad theoretical and experimental development \cite{Dodonov2010,Dodonov2020,Nation2012}. 
Rapidly changing boundary conditions may be implemented in superconducting circuits, with proposals and calculations of their radiation \cite{Johansson2009,Johansson2010} and experimental observations of vacuum emission \cite{Wilson2011,Lahteenmaki2013}. These systems motivate comparing particle multiplicity with energy.

We focus on asymptotically resting mirrors, which start and end at zero speed and pass through one or several velocity peaks.
There are no horizons, and the incoming and outgoing scattering problems are complete on both sides of the mirror (recall that we work in 1+1 dimensions). Asymptotic rest does not, by itself, imply that a mirror moves only a finite distance; if the speed decays sufficiently slowly, the mirror may come from one spatial infinity and then escape to the same or the other infinity.
These possibilities let us ask which features of the radiation are local to the fast part of the motion and which depend on the complete trajectory.

Examples of finite particle production for smooth asymptotically resting motion are already known in the literature \cite{Good:2020rmk}. See the early example by Walker-Davies \cite{WD1982}. See also Arctx trajectories \cite{GAE2013} and the betaK (Leonardo da Vinci) mirror \cite{FiniteEntropy,Good:2023ncu}. In those works, the total particle counts were evaluated only numerically. 
In the non-relativistic approximation, certain types of motion can be treated analytically (with Fourier transforms); see \cite{RoundTrip} for finite particle production from slow round trips. 
Our goal here is more ambitious: we wish to analytically complete both number and energy calculations for a family with fully relativistic velocity dependence, then use the family to test a general prediction for an ultrarelativistic scaling law: under certain assumptions, the total particle count grows linearly with maximal rapidity, while the total energy depends on the shape of the rapidity peak.

Our first example is an exceptionally analytic polynomial family of asymptotically resting trajectories whose position monotonically travels from one spatial infinity to the other. We begin with its simplest member in Sec.~\ref{sec:cubic} and then discuss the full family in Sec.~\ref{sec:family}. We then explore analytically interesting non-monotonic mirror motions. Section~\ref{sec:returns} solves returning (round-trip) motions.
Section~\ref{sec:local} explains how the motion near speed maxima determines the leading ultrarelativistic scaling of the total particle number and radiated energy. Appendices cover technical derivations of some of the results presented in the main part.

\section{Cubic flight}
\label{sec:cubic}

We start with a simple example that illustrates the main features we are after: asymptotically resting motion with one rapidity peak, and its exact particle count and total energy counts. In this section we also set our conventions.

\subsection{The motion}

Our setup is 1+1d QFT on a line, one real scalar field. 
The field obeys $(\partial_t^2-\partial_x^2)\phi=0$ and the Dirichlet condition $\phi |_\text{mirror}=0$ \cite{FD1976}.
We set a perfectly reflecting boundary (mirror) at a point $z$ which moves with time $t$.
Related descriptions use an effective Hamiltonian for a moving cavity wall or changing dielectric medium \cite{Law1994}, and resonant enhancement in a vibrating cavity \cite{Lambrecht1996}. 

We take the mirror to move along a specific trajectory, which, in the case of the cubic flight, is given by\footnote{The cubic flight is the component of a Newtonian parabolic Kepler orbit perpendicular to the parabola's symmetry axis.}
\begin{equation}
    t(z)=-\frac{z}{V}-\frac{z^3}{3V^3\tau^2}.
\label{eq:cubic}
\end{equation}
Here and below, we denote the maximal speed of the mirror by $V$, $0< V < 1$, while $\tau$ is a free parameter that sets a time scale.
We express the trajectory through $t(z)$ rather than $z(t)$ because the resulting formula is simpler. Since the motion is monotonic, there is no ambiguity in passing between these two descriptions.
As $t$ runs from $-\infty$ to $+\infty$, the mirror travels monotonically from $z=+\infty$ to $z=-\infty$, with the speed vanishing asymptotically at both ends. 

The instantaneous velocity is
\begin{equation}
 v(z)=\frac{1}{t'(z)}=-\frac{V}{1+z^2/(V\tau)^2}.
 \label{eq:cubicspeed}
\end{equation}
With the maximal speed being $V$, we denote the maximal rapidity and Lorentz factor by
\begin{equation}
 \eta_* =\atanh V,\qquad \gamma=\cosh\eta_*=(1-V^2)^{-1/2}.
 \label{eq:peak}
\end{equation}
That maximum is achieved at $z=0$, $t=0$, with quadratic dependence near the extremum:
\begin{equation}
 |v(t)| = V - V(t/\tau)^2+O((t/\tau)^4).
 \label{eq:cubicdefect}
\end{equation}
This formula will be important later.

The trajectory $t(z)$ can also be inverted and written in the form $z(t)$
\begin{equation}
z(t)=-2V\tau\,\sinh\!\left[
\frac13\operatorname{arsinh}\!\left(\frac{3t}{2\tau}\right)
\right].
\label{eq:cubic_z_of_t}
\end{equation}
This will be used for a comparison later.

\subsection{Total particles and energy}

The creation of quanta by the moving mirror can be described with the help of the Bogoliubov coefficients. Because we are working on a spatial line, there are in principle two sets of coefficients, one for each side of the mirror.
For the right-hand side, we have \cite{WalkerEnergy,GAE2013}
\begin{equation}
 \beta^R_{\omega\omega'}=
 \frac{\sqrt{\omega\omega'}}{\pi(\omega+\omega')}
 \int_\R v(t)e^{-i(\omega+\omega')t+i(\omega-\omega')z(t)}\dd t .
 \label{eq:beta}
\end{equation}
Here $\omega>0$ is an outgoing frequency and $\omega'>0$ an incoming frequency \cite{GAE2013}. 

In the incoming vacuum, the outgoing spectral density is obtained from the squared Bogoliubov coefficient \cite{WalkerEnergy,Ritus1998}:
\begin{equation}
 n_R(\omega)=\int_0^\infty|\beta^R_{\omega\omega'}|^2\dd\omega',
 \label{eq:occupation}
\end{equation}
and analogously for $n_L$. Integrating the spectral density gives the expected particle number, while weighting it by the outgoing frequency gives the expected energy \cite{WalkerEnergy,Ritus1998}. Summing the two exterior sides, we define
\begin{align}
 N&=\int_0^\infty[n_R(\omega)+n_L(\omega)]\dd\omega,\nonumber\\
 E&=\int_0^\infty\omega[n_R(\omega)+n_L(\omega)]\dd\omega,
 \qquad \bar\omega=E/N.
 \label{eq:observables}
\end{align}
For radiation on the left, one can use the identity $|\beta^L_{\omega\omega'}|=|\beta^R_{\omega'\omega}|$ \cite{Ritus1998}. Consequently, for the total particle counts it follows that $N_L=N_R=N/2$, although the spectra and energies on the two sides need not coincide \cite{Ritus1998}. We define $\bar\omega$ as the ratio of ensemble-averaged energy and number, for finite $E$ and $0<N<\infty$.

For the cubic flight trajectory Eq.~\eqref{eq:cubic}, the beta coefficients are expressible through the Airy function:
\begin{equation}
 \beta^R_{\omega\omega'}=-V\tau\sqrt{1-\mu^2}\,(\tau\Omega)^{-1/3}
 \Ai\bigl[(1+V\mu)(\tau\Omega)^{2/3}\bigr].
 \label{eq:cubicbeta}
\end{equation}
Here we introduce the following shorthand notation for the frequency sum and frequency asymmetry:
\begin{equation}
 \Omega=\omega+\omega',\qquad
 \mu=\frac{\omega-\omega'}{\omega+\omega'}.
 \label{eq:frequency}
\end{equation}
The total particle count can be computed exactly by changing the integration variables from the frequencies $\omega$, $\omega'$ to $\Omega$ and $\mu$.
The Jacobian is $\dd\omega\dd\omega'=\frac12\Omega\dd\Omega\dd\mu$, with $\Omega>0$ and $-1<\mu<1$. 
Using the integral identity \cite{DLMF}
\begin{equation}
 J_3=\int_0^\infty x\Ai(x)^2\dd x=\frac{1}{6\pi\sqrt3}.
 \label{eq:airy}
\end{equation}
we arrive at
\begin{equation}
 N_1=\frac32 J_3 V^2\int_{-1}^{1}
 \frac{1-\mu^2}{(1+V\mu)^2}\dd\mu=C_1F(\eta_*),
 \label{eq:cubicN}
\end{equation}
where the constant $C_1$ and the function $F$ are given by explicit formulas:
\begin{equation}
 C_1=\frac{1}{\pi\sqrt3},\qquad
 F(\eta)=\eta\coth\eta-1,\qquad F(0)=0.
 \label{eq:F}
\end{equation}
The motivation for introducing the subscript 1 will be clear in the next section.
Note that the exact number does not depend on the time scale parameter $\tau$ (but the betas \eqref{eq:cubicbeta} and the spectrum certainly do depend on it).

The total energy can be easily found from the stress tensor calculation \cite{FD1976}, which gives a formula analogous to the relativistic Larmor formula \cite{QuantumPower}.
Technical details of the derivation are not important here; we refer the interested reader to Appendix~\ref{app:flightenergy}.
The answer for the cubic flight (both sides) is
\begin{equation}
E_1(V)=
\frac{V^2\left(2+3\sqrt{1-V^2}\right)}
{48\sqrt{2}\,\tau
\left[(1-V^2)\left(1+\sqrt{1-V^2}\right)\right]^{3/2}}.
 \label{eq:cubicE}
\end{equation}

Finally, let us discuss the scaling in the ultrarelativistic limit. 
As $V\to 1$, 
\begin{equation}
 N_1 \approx C_1\log\gamma \approx C_1 \eta_*,\qquad
 E_1\sim\frac{\gamma^3}{24\sqrt2\tau}.
 \label{eq:cubiclimit}
\end{equation}
Thus $\bar\omega_1\sim\gamma^3/(24\sqrt2\tau C_1\log\gamma)$. The cubic growth of energy is much faster than the logarithmic growth of the particle number; i.e., the radiation chiefly hardens.

\section{Polynomial flights of arbitrary fixed order}
\label{sec:family}

In this section we generalize the cubic flight trajectory from Eq.~\eqref{eq:cubic} and discuss the particle and energy scaling.

\subsection{The trajectory and its exact particle count}

Consider the following polynomial generalization of the cubic flight trajectory:
\begin{equation}
 \begin{gathered}
 t(z)=-\frac{z}{V}\left[1+\frac{(z/V\tau)^{2m}}{2m+1}\right],\\
 v(z)=-\frac{V}{1+(z/V\tau)^{2m}}.
 \end{gathered}
 \label{eq:family}
\end{equation}
As before, $V$ is the maximal speed, and $\tau$ is an arbitrary parameter setting the time scale.
The new parameter $m$ is an arbitrary positive integer. The trajectory Eq.~\eqref{eq:cubic} corresponds to $m=1$; hence the subscript 1 in Eqs.~\eqref{eq:cubicN}-\eqref{eq:cubiclimit}.
For a general trajectory in this polynomial family, there is no simple $z(t)$ expression like in Eq.~\eqref{eq:cubic_z_of_t}.

\begin{figure*}[t]
\centering
\includegraphics[width=.98\textwidth]{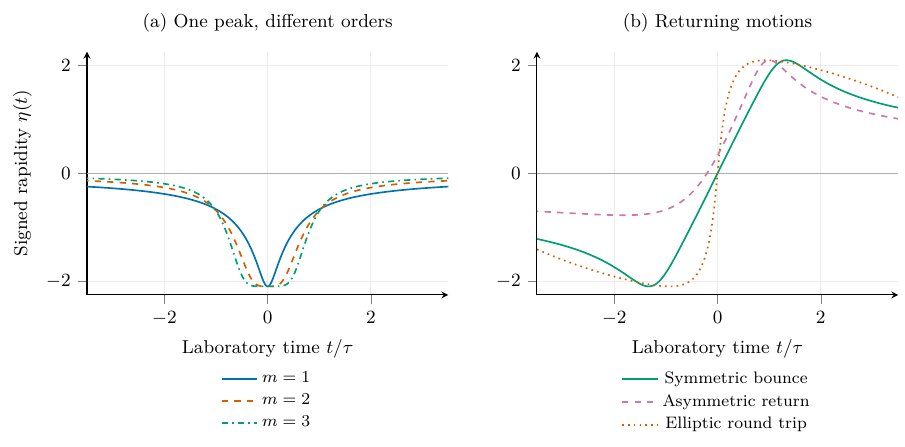}
\caption{
Signed rapidity as a function of laboratory time. (a) The polynomial trajectories in Eq.~\eqref{eq:family} have $m=1,2,3$, common maximum speed $V=0.97$, and common $\tau$. They have equal maximum rapidity magnitude, but $V-|v(t)|$ grows as $t^{2m}$ near the maximum speed. (b) The symmetric bounce, Eq.~\eqref{eq:bounceexplicit}, has $V=0.97$; the asymmetric return, Eq.~\eqref{eq:clockexplicit}, has $V_+=0.97,V_-=0.65$; and the elliptic round trip, Eq.~\eqref{eq:pendtraj}, has actual maximum speed $0.97$, with $L/\tau\approx5.18$. The same $\tau$ is used throughout. Time varies along each curve while its parameters remain fixed. The elliptic mirror turns at $t=z=0$ and tends to $z=L$ at both temporal ends. The finite time window displays the rapid motion; all curves tend to zero rapidity as $t\to\pm\infty$.
}
\label{fig:rapidity}
\end{figure*}

The mirror runs monotonically from $z = + \infty$ to $z= - \infty$. The speed vanishes in the asymptotic past and future, reaching its maximum at $t=z=0$, where the velocity is negative. Near this point,
\begin{equation}
 |v(t)|= V - V(t/\tau)^{2m}
 +O((t/\tau)^{4m}).
 \label{eq:familydefect}
\end{equation}
It's easy to see that the parameter $m$ controls the order of the extremum: quadratic, quartic, and so on.
Fig.~\ref{fig:rapidity}(a) illustrates the behavior.

The beta Bogoliubov coefficients can be expressed through the generalized Airy function. We use the following notation for this function and one of its moments:\footnote{For real $x$, the first integral is equivalently $\pi^{-1}\int_0^\infty\cos[s^{2m+1}/(2m+1)+xs]\dd s$, interpreted as an oscillatory integral. This is the generalized Airy function defined in Ref.~\cite{HigherAiry}, whose index is $m$ rather than our $2m+1$. In particular, $\A_3=\Ai$.}
\begin{align}
 \A_{2m+1}(x)&=\frac{1}{2\pi}\int_\R
 e^{i[s^{2m+1}/(2m+1)+xs]}\dd s,\nonumber\\
 J_{2m+1}&=\int_0^\infty x\A_{2m+1}(x)^2\dd x.
 \label{eq:airyfamily}
\end{align}
In terms of $\A_{2m+1}$, the Bogoliubov coefficients are given by
\begin{align}
 \beta^R_{\omega\omega'}={}&-V\tau\sqrt{1-\mu^2}\,(\tau\Omega)^{-1/(2m+1)}\nonumber\\
 &\times\A_{2m+1}\bigl[(1+V\mu)(\tau\Omega)^{2m/(2m+1)}\bigr].
 \label{eq:familybeta}
\end{align}
Here $\Omega$ and $\mu$ denote the frequency sum and asymmetry introduced in Eq.~\eqref{eq:frequency}.
In terms of these variables, the total particle integral can be solved again. 
The result is proportional to the same function $F$ from Eq.~\eqref{eq:F}:
\begin{equation}
N_m = C_m F(\eta_*), \qquad C_m=\frac{2(2m+1)}{m}J_{2m+1}.
 \label{eq:familyN}
\end{equation}
The integral $J_{2m+1}$ is evaluated in Appendix~\ref{app:exact}, giving a finite elementary sum
\begin{equation}
 C_m=\frac{1}{\pi m}\sum_{p=1}^{m}
       \cot\!\left(\frac{\pi p}{2m+1}\right).
 \label{eq:Cm}
\end{equation}
Note that every term in this sum is positive, so that the whole expression is also manifestly positive.
The first values are
\begin{equation}
 \begin{gathered}
 C_2=\frac1\pi\sqrt{\frac{5+\sqrt5}{10}},\\
 C_3\simeq0.329158,\qquad C_4\simeq0.373450.
 \end{gathered}
 \label{eq:constants}
\end{equation}
Note that $C_m > 0$ always, as follows from the $x\A_{2m+1}(x)^2$ piece in Eq.~\eqref{eq:airyfamily}.

\subsection{Elementary total energy and relativistic scaling}

The total energy is also elementary for every $m$.
The computation itself is a bit technical, but the result is relatively simple:
\begin{align}
 E_m={}&\frac{1- 1/(2m)}{48\tau\sin[\pi/(2m)]}
 \biggl[(1-V)^{-1-1/(2m)}\nonumber\\
 &+(1+V)^{-1-1/(2m)}\nonumber\\
 &+\frac{2m}{V}\bigl((1+V)^{-1/(2m)}-(1-V)^{-1/(2m)}\bigr)\biggr].
 \label{eq:familyE}
\end{align}
We refer to Appendix~\ref{app:flightenergy} for the details.
The apparent singularity at $V=0$ is removable, with $E_m(0)=0$. Evaluation near zero should use a series or sufficient precision to resolve the cancellation.
For $m=1$, we recover Eq.~\eqref{eq:cubicE}.

In the ultrarelativistic limit $V \to 1$, the scaling of the total particles and energy can be easily derived from Eqs.~\eqref{eq:familyN} and \eqref{eq:familyE}:
\begin{align}
 N_m&\approx C_m\log\gamma \approx C_m \eta_* , \nonumber\\
 E_m&\approx \frac{D_m}{\tau}\gamma^{2+1/m},\qquad
 D_m=\frac{2^{1/(2m)}(1-1/(2m))}{24\sin[\pi/(2m)]}.
 \label{eq:familylimit}
\end{align}
Here, $\gamma$ is the Lorentz factor at the peak. 
For large $m$ in the ultrarelativistic regime, the mean
energy per quantum obeys
\begin{equation}
\bar\omega_m\sim \frac{ \pi m }{ 24 \tau \log m}
\frac{\gamma^{2+1/m}}{\log\gamma }.
\label{eq:mean}
\end{equation}
As one can see, the energy per quantum grows with $m$ and with $\gamma$ (for sufficiently large $m$).

\section{Returning motions and finite-speed comparisons}
\label{sec:returns}

Of course, peak order is not the only way to vary a trajectory. We next keep quadratic extrema while changing their number and rapidities, and we demand a return-trip motion. These comparisons separate local shape from the trajectory's global properties.

\subsection{Symmetric bounce and the two-parameter cubic family}

In this subsection we focus on the following trajectory:\footnote{\label{ft:bounce_param_s}
Secretly, this trajectory is related to the cubic flight trajectory. One can parametrize it as
$t(s)=\tau(s+s^3/3)$, $z(s)=V\tau s^2$ with an auxiliary parameter $s$; one can compare to Eq.~\eqref{eq:cubic}.
}

\begin{equation}
 z(t)=4V\tau\sinh^2\!\left[\frac13\operatorname{arsinh}\!\left(\frac{3t}{2\tau}\right)\right].
 \label{eq:bounceexplicit}
\end{equation}
As before, $V$ is the maximum speed, and the parameter $\tau$ sets the time scale.
The mirror starts at $z=+\infty$ with $\dot{z}\to0$ as $t\to-\infty$, reaches $z=0$ and turns at $t=0$, and returns to $z=+\infty$ with $\dot{z}\to0$ as $t\to+\infty$. It has two speed maxima, at velocities $\dot{z} = \pm V$.

For this trajectory\footnote{This symmetric bounce can be obtained from the component of a Newtonian parabolic Kepler orbit parallel to the parabola's symmetry axis.}, one can compute the analytic (and elementary) total particle count and energy.
A more general consideration presented in Appendix~\ref{app:clock} gives
\begin{align}
 N_{\rm b}&=C_1[3F(2\eta_*)-4F(\eta_*)],\nonumber\\
 E_{\rm b}&=  E_{\rm b}(V)=\frac{V^2(7-6V^2)}{48\tau(1-V^2)^{3/2}},   
 \label{eq:bouncetotals}
\end{align}
with the same function $F$ defined in Eq.~\eqref{eq:F}.
In the relativistic limit $V \to 1$, we have
\begin{equation}
    N_{\rm b} \approx 2C_1\log\gamma , \quad
    E_{\rm b} \approx \gamma^3/(48\tau),
\label{sym_b_totals_asympt}
\end{equation}
where $C_1 = 1/(\pi \sqrt{3})$ is defined in Eq.~\eqref{eq:F}.
Comparing with Eq.~\eqref{eq:cubiclimit}, one can see that the presence of two velocity peaks doubles the leading particle count coefficient. The energy coefficient does not have to be the same, as it depends on the time rescaling parameter $\tau$.

The bounce trajectory \eqref{eq:bounceexplicit} is again a member of a larger family.
Let us introduce two speed parameters $V_+$, $V_-$ such that $0\le V_+,V_-<1$ and $V_++V_->0$. 
The trajectory is then explicitly given by
\begin{align}
 z(t)={}&2\tau(V_+-V_-)\sinh\!\left[\frac13\operatorname{arsinh}\!\left(\frac{3t}{2\tau}\right)\right]\nonumber\\
 &+4\tau\sqrt{V_+ V_-}\sinh^2\!\left[\frac13\operatorname{arsinh}\!\left(\frac{3t}{2\tau}\right)\right].
 \label{eq:clockexplicit}
\end{align}
For $V_+ = V_- = V$ this reduces back to Eq.~\eqref{eq:bounceexplicit}\footnote{
In terms of the parameter $s$ from footnote~\ref{ft:bounce_param_s}, this trajectory can be written as $t(s)=\tau(s+s^3/3)$, $z(s)=\tau[(V_+-V_-)s+\sqrt{V_+ V_-}\,s^2]$.
}, while $V_+=0$ or $V_-=0$ are versions of the cubic flight, cf. Eq.~\eqref{eq:cubic_z_of_t}.
For $V_+ V_->0$, the mirror begins at $z=+\infty$ with $v\to0^-$, turns once at the finite minimum $z=-\tau(V_+-V_-)^2/(4\sqrt{V_+ V_-})$, and returns to $z=+\infty$ with $v\to0^+$. Its maximum and minimum velocities are $+V_+$ and $-V_-$. When $V_+=0$, it instead moves monotonically from $+\infty$ to $-\infty$; when $V_-=0$, it moves from $-\infty$ to $+\infty$. Both limiting flights have zero asymptotic velocity.
The corresponding fixed rapidity magnitudes and Lorentz factors are
\begin{equation}
 \begin{gathered}
 \eta_+=\atanh V_+,\qquad \eta_-=\atanh V_-,\\
 \gamma_+=\cosh\eta_+,\qquad \gamma_-=\cosh\eta_-.
 \end{gathered}
 \label{eq:clockpeaks}
\end{equation}

The total particle count can be written in terms of the function $F$ from Eq.~\eqref{eq:F}
\begin{equation}
 N_{\rm c}=C_1[3F(\eta_++\eta_-)-2F(\eta_+)-2F(\eta_-)].
 \label{eq:clockN}
\end{equation}
See Appendix~\ref{app:clock} for details. The negative coefficients in this formula arise from an algebraic decomposition of a positive norm; they are not negative particle populations. The total energy can also be computed, but the resulting expression is a bit more involved and is also presented in Appendix~\ref{app:clock}.

In the relativistic limit, when at least one of the speeds $V_+$, $V_-$ approaches the speed of light, the leading particle count is given by a simple sum
\begin{equation}
    N_c \approx C_1 (\log\gamma_+ + \log\gamma_-) \approx C_1 ( \eta_+ + \eta_-)
\label{asymmetric_bounce_rel_particles}
\end{equation}
For $V_+ = V_- = V$ this reduces to Eq.~\eqref{sym_b_totals_asympt}, while at $V_+=0$ or $V_-=0$ we recover Eq.~\eqref{eq:cubiclimit}. As one can see, at the particle-count level there is no interference between the two peaks. There is no ambiguity, as the leading particle count depends only on the peak's height.

The leading behavior of the total energy is given by
\begin{equation}
E_{\rm c}\sim
\frac{V_+^2\gamma_+^3+V_-^2\gamma_-^3}
{24\sqrt2\,\tau(V_++V_-)^{3/2}}
\label{asymmetric_bounce_rel_energy}
\end{equation}
One should not be surprised that this formula does not appear to be a simple sum.
The coefficient in the energy asymptotics depends not only on the peak's height, but also on its shape; hence the explicit dependence on  $\tau$, $V_+$ and $V_-$ in the denominator of Eq.~\eqref{asymmetric_bounce_rel_energy}. By rescaling $\tau$ to absorb $(V_++V_-)^{3/2}$, we can again bring the energy formula to a simple sum. There is no interference between the two peaks, once their shapes (the quadratic coefficient in the Taylor series expansion) are properly taken into account.
If only one of the speeds, either $V_+$ or $V_-$, goes ultra-relativistic, the contribution from the second peak becomes subleading and can be ignored at leading order.

\begin{figure*}[t]
\centering
\includegraphics[width=.98\textwidth]{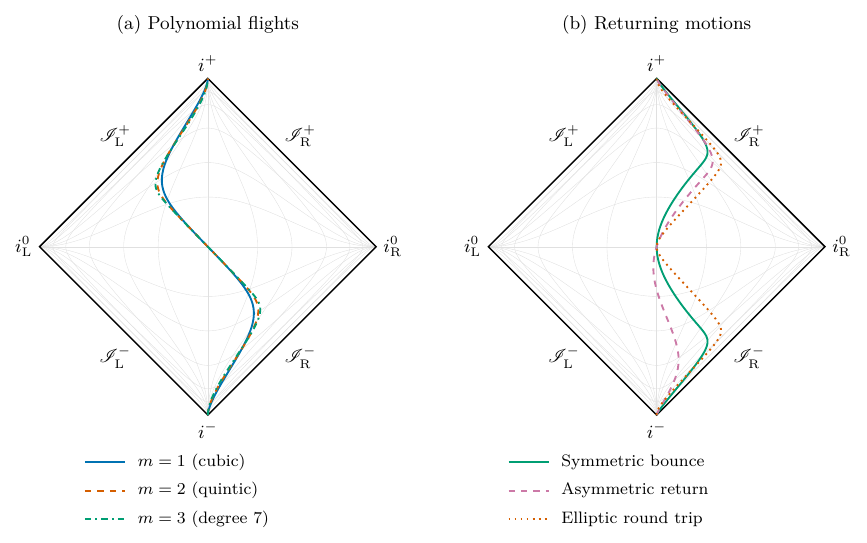}
\caption{
Penrose diagrams of the studied trajectories. (a) Polynomial flights, Eq.~\eqref{eq:family}, with $m=1,2,3$ and common fixed $V=0.97,\tau$; cubic flight is $m=1$. (b) Symmetric bounce at $V=0.97$, two-parameter cubic motion at $V_+=0.97,V_-=0.65$, and the elliptic round trip at maximum speed $0.97$, with the same $\tau$. Their trajectories are Eqs.~\eqref{eq:bounceexplicit}, \eqref{eq:clockexplicit}, and \eqref{eq:pendtraj}, respectively, with the same parameters as in Fig.~\ref{fig:rapidity}. The elliptic mirror turns at $t=z=0$ and approaches the same finite position $z=L$ in the asymptotic past and future. Time runs along each curve while its parameters remain fixed. The diamond edges are left/right future and past null infinity, $\mathscr I_{\mathrm L,\mathrm R}^{\pm}$; its side corners are spatial infinity $i^0_{\mathrm L,\mathrm R}$. All curves terminate at $i^\pm$. Thin gray curves are lines of constant laboratory time or position, not additional trajectories. The plots preserve the positions and time origins of the defining equations.
}
\label{fig:penrose}
\end{figure*}

Table~\ref{tab:trajectories} collects the number and energy properties of these trajectories. It distinguishes exact finite-speed evaluation from a leading asymptotic law.

\begin{table*}[t]
\caption{Exact particle counts and energies for the trajectories in Eqs.~\eqref{eq:cubic}, \eqref{eq:family}, \eqref{eq:bounceexplicit}, \eqref{eq:clockexplicit}, and \eqref{eq:pendtraj}. All totals include both sides. ``Elementary'' and ``elliptic'' specify the form of the completed finite-speed results. The leading terms apply as the indicated speed maxima tend to unity, with $\tau$ fixed and, for the polynomial family, fixed $m$. In the two-parameter row, $\gamma_j$ denotes each diverging peak Lorentz factor. The coefficients $C_m,D_m$ are given in Eqs.~\eqref{eq:Cm} and \eqref{eq:familylimit}. Quadratic and order-$2m$ maxima refer to the leading decrease of speed away from its maximum. The elliptic row concerns $L/\tau\to\infty$; its quadratic peaks flatten in this limit, so the uniform hypotheses of the local theorem do not apply. Its totals and limits are Eqs.~\eqref{eq:pendtotals} and \eqref{eq:pendlimit}. The polynomial family includes cubic and quintic flight; the two-parameter family includes flight and bounce.}
\label{tab:trajectories}
\begin{ruledtabular}
\begin{tabular}{p{.19\textwidth}p{.23\textwidth}p{.22\textwidth}p{.23\textwidth}}
Trajectory & Endpoints; fast events & Particle count: exact status; leading term & Energy: exact status; leading $\tau E$\\
\hline
\tcell{Cubic flight; Eq.~\eqref{eq:cubic}} & \tcell{Opposite infinities; one quadratic} & \tcell{Elementary; $C_1\log\gamma$} & \tcell{Elementary; $\gamma^3/(24\sqrt2)$}\\
\tcell{Polynomial flight, $m\ge1$; Eq.~\eqref{eq:family}} & \tcell{Opposite infinities; one of order $2m$} & \tcell{Elementary; $C_m\log\gamma$} & \tcell{Elementary; $D_m\gamma^{2+1/m}$}\\
\tcell{Symmetric bounce; Eq.~\eqref{eq:bounceexplicit}} & \tcell{Same infinity; two quadratic} & \tcell{Elementary; $2C_1\log\gamma$} & \tcell{Elementary; $\gamma^3/48$}\\
\tcell{Two-parameter cubic motion; Eq.~\eqref{eq:clockexplicit}} & \tcell{Same infinity for $V_+V_->0$; up to two quadratic} & \tcell{Elementary; $C_1\sum_{\rm fast}\log\gamma_j$} & \tcell{Elementary; Eq.~\eqref{asymmetric_bounce_rel_energy}}\\
\tcell{Elliptic round trip; Eq.~\eqref{eq:pendtraj}} & \tcell{Same finite position $L$; two quadratic, flattening peaks} & \tcell{Elliptic; $\dfrac4{\pi^2}(\log\gamma)^2$} & \tcell{Elementary; $\gamma^2/8$}\\
\end{tabular}
\end{ruledtabular}
\end{table*}

\section{Local laws for number and energy}
\label{sec:local}

For the polynomial flights and cubic returning motions above, the leading ultrarelativistic particle count grows linearly with the peak rapidities, while the total radiated energy depends on the shapes of the speed peaks.
In this section we will discuss how general this result is. The elliptic round trip from Appendix~\ref{app:pendulum} lies outside this uniform class, as explained below. 
We present here the main points of the argument, referring to \cite{MirrorCountsGitHub} for a more rigorous proof.

Consider a mirror on a generic smooth trajectory. Assume that its speed (absolute value of the velocity) vanishes in the asymptotic past and future, while going through a finite number of local maxima at intermediate times $t_j$. 
Our idea is to derive the leading contributions from each of the peaks and then compare them.

Let us assume that there is a unique maximum at some $t_0$:
\begin{equation}
     v(t_0 + \Delta t)  \approx V - V (\Delta t / \tau)^{2 m}
\label{velocity_expansion_general}
\end{equation}
Here, $0<V<1$ is the maximal speed, $m$ is a positive integer, and $\tau$ effectively sets the time scale of the fast-speed event. Without loss of generality, we can set $t_0 = 0$. The case with negative velocity is completely analogous.

For several separated fast events, one should sum the local contributions coming from every near-lightspeed peak, even when those limits occur
at unequal rates.

\subsection{Leading radiated energy}

The leading energy behavior is relatively simple to derive, so let us start with that.
We can directly use the stress tensor calculation result \cite{FD1976,QuantumPower}:
\begin{equation}
 E=\frac{1}{6\pi}\int_\R\frac{\dot v^2}{(1-v^2)^3}\dd t.
 \label{eq:stress}
\end{equation}
Assume the relativistic limit $V \to 1$. Near this maximum, the denominator in \eqref{eq:stress} becomes small and wins over $\dot{v}$ vanishing at the extremum. The leading contribution comes from that event.
Plugging in the expansion \eqref{velocity_expansion_general} and using $1-V^2 \approx 2(1-V)$, we obtain for the corresponding contribution
\begin{equation}
    E \approx \frac{1}{6\pi} \int \frac{ 4 m^2 V^2 t^{4m - 2}/\tau^{4m}}{ 8 (1 - V + V (t/\tau)^{2m})^3 } \dd t.
\end{equation}
By rescaling the integration variable as $t \to ((1-V)/V)^{1/(2m)} \tau \tilde{t}$, we immediately obtain
\begin{equation}
    E \sim \frac{ 1 }{\tau (1-V)^{1 + 1/(2m)}}.
\label{energy_asymptotic_main}
\end{equation}
A more careful analysis gives $\tau E\sim D_m\gamma^{2+1/m}$, with $D_m$ as in Eq.~\eqref{eq:familylimit}. See Ref.~\cite{MirrorCountsGitHub} for details. 

The same result can be derived by a slightly different scaling argument, which will turn out to be useful for the particle count.
The time scale relevant for the fast-speed event is determined by the time scale over which the denominator in Eq.~\eqref{eq:stress} is small; with Eq.~\eqref{velocity_expansion_general} this implies
\begin{equation}
    1-V \sim V (\Delta t / \tau)^{2 m}
    \Rightarrow 
    \Delta t \sim \tau (1-V)^{1/(2m)}.
\end{equation}
Over that time scale, we have
\begin{equation}
\begin{aligned}
    \dot{v} &\sim \left( \frac{\Delta t}{\tau} \right)^{2m} \frac{1}{\Delta t} \sim \frac{1}{\tau} (1-V)^{1 - 1/(2m)}, \\
    1-v^2 &\sim 1-V, \\
    dt &\sim \Delta t \sim \tau (1-V)^{1/(2m)}.
\end{aligned}
\label{time_scale_estimates}
\end{equation}
Plugging these estimates into Eq.~\eqref{eq:stress} we again recover Eq.~\eqref{energy_asymptotic_main}.

Explicit $\tau$-dependence of the energy in  Eq.~\eqref{energy_asymptotic_main} signals the fact that the energy output depends on the shape (e.g. width) of the velocity peak.
In the ultrarelativistic limit this clearly agrees with our observations above, see e.g. Eq.~\eqref{eq:familylimit}.
The coefficient in Eq.~\eqref{energy_asymptotic_main} reduces to a definite beta integral and is evaluated in the companion theorem proof~\cite{MirrorCountsGitHub}.

\subsection{Particle count formula}

In order to relate the velocity expansion \eqref{velocity_expansion_general} to the particle count, we need to rewrite the counting formula from Eq.~\eqref{eq:observables} in a more suitable way.
By passing from the integration over frequencies to the frequency sum $\Omega$ and asymmetry $\mu$ defined in Eq.~\eqref{eq:frequency} and then using Parseval's identity, it is possible to reduce the number of integrations by one and pass to the following representation:
\begin{equation}
 \begin{aligned}
 N={}&\frac{1}{8\pi^2}\int_{-1}^{1}(1-\mu^2)\,\mathrm{d}\mu\\
 &\times\iint_{\mathbb R^2}
 \frac{\left[\dfrac{v(t)-v(t')}{t-t'}\right]^2
 \,\mathrm{d}t\,\mathrm{d}t'}
 {[1-\mu v(t)][1-\mu v(t')][1-\mu\bar v(t,t')]^2}.
 \end{aligned}
 \label{eq:pn-positive-number}
\end{equation}
Here, $\bar v(t,t')$ is the average velocity,
\begin{equation}
 \bar v(t,t')=\frac{z(t)-z(t')}{t-t'}.
 \label{eq:pn-chord-velocity}
\end{equation}
Details of the derivation are presented in Appendix~\ref{app:positive}.
Note that $\mu$ is integrated over in this expression.

To get the leading particle count behavior, we can again use the time scale estimates.
Because the small denominators in Eq.~\eqref{eq:pn-positive-number} involve $\mu$, the relevant time scale also contains it:
\begin{equation}
    \Delta t \sim \tau (1 - \mu V)^{1/(2m)}
\end{equation}
Then, using estimates similar to those in \eqref{time_scale_estimates}, we conclude that the integral in Eq.~\eqref{eq:pn-positive-number} reduces to
\begin{equation}
    N \sim \int_{-1}^{1} \frac{ 1-\mu^2 }{ (1 - \mu V)^2 } \dd\mu
\end{equation}
As $V \to 1$, the leading contribution to this integral comes from the vicinity of the endpoint $\mu \to 1$, with the answer
\begin{equation}
    N \sim - \log (1 - V) 
\end{equation}
The minus sign makes the expression positive.
This reproduces our observations above. 
Again, the coefficient in front of the log here can also be computed in closed form, and it matches Eqs.~\eqref{eq:Cm} and \eqref{eq:familylimit}, see \cite{MirrorCountsGitHub}.

To end this section, let us make a remark on what other assumptions implicitly enter into this scaling result. Let's say that we have a trajectory that depends on some parameter, and suppose that by tuning that parameter we can go to the ultrarelativistic limit. If a trajectory is smooth and has a finite number of rapidity peaks at every given value of that parameter, but not uniformly through the entire range of that parameter, then the scaling result presented here may fail. For example, the limiting trajectory may lose smoothness, or develop a finite region where its speed goes to the speed of light.
The elliptic round trip in Appendix~\ref{app:pendulum} illustrates this limitation: its count grows as $N\propto(\log\gamma)^2$, or equivalently as the square of the peak rapidity. Its speed peaks flatten in the ultrarelativistic limit, placing it outside the uniform assumptions of the logarithmic scaling result.

\section{Discussion}
\label{sec:discussion}

The exact totals distinguish changes in particle multiplicity from changes in energy. For cubic and quintic flight at common peak speed and time scale, $N_2/N_1=C_2/C_1>1$, whereas $E_2/E_1\sim(D_2/D_1)\gamma^{-1/2}\to0$. Thus sufficiently relativistic quintic flight produces more quanta but less energy. This result, made possible by the completed particle-number integral with full velocity dependence, complements the already existing analytic spectra with numerically evaluated totals~\cite{FiniteEntropy,Good:2020rmk} and nonrelativistic round-trip counts~\cite{RoundTrip}.

Under the theorem's assumptions, the main growth of particle
number and energy comes from the fastest parts of the motion.
The particle count depends on the peak speeds \textit{and} the shape of the peaks; the energy also depends on the peak width in time. Contributions from separate peaks add at leading order.
Slower motion and interference can still affect the finer
details of the radiation.

Even a small change to a flat speed peak can change the leading particle count. For example, adding a cubic term to quintic flight ($m=2$ in Eq.~\eqref{eq:family}) gives
\begin{equation}
\begin{aligned}
 t(s)&=\tau\left(s+\frac{\varepsilon s^3}{3}+\frac{s^5}{5}\right),\\
 z(s)&=-V\tau s,
\end{aligned}
\label{eq:unfolding}
\end{equation}
where $\varepsilon\ge0$ controls the perturbation and $0<V<1$ is the maximum speed. The velocity is $v(s)=-V/(1+\varepsilon s^2+s^4)$, and the motion remains smooth, subluminal, and asymptotically at rest. At $\varepsilon=0$, the speed peak is quartic and the leading count coefficient is $C_2$. Any fixed $\varepsilon>0$, however small, makes the peak quadratic and gives $C_1$ as $V\to1$. If $\varepsilon\to0$ at the same time, the ratio $\varepsilon/\sqrt{1-V}$ measures the competition between the two terms near the peak; this transition needs analysis beyond the fixed-order theorem. Peaks that merge, change order, or have vanishing coefficients likewise need separate treatment. The elliptic round trip illustrates the broader caution: smooth motion that starts and ends at rest does not, by itself, guarantee logarithmic growth of the particle count.

Overall, we also caution that our results describe a prescribed ideal Dirichlet boundary and do not determine the external drive, recoil, or backreaction. For example, backreaction and open-system models address these dynamics~\cite{Hotta1994,KumarBackreaction,ButeraNoise}, while dispersive mirrors include boundary response and radiation reaction~\cite{Barton1995,Calogeracos1995}, and scattering approaches to dissipation allow high-frequency transparency~\cite{Jaekel1992}. Partial reflection and semitransparency provide further extensions~\cite{Obadia2001,Nicolaevici2009}.

Finite transverse size and thickness affect relativistic semitransparent-mirror radiation~\cite{LinChenFinite}. Higher-dimensional Dirichlet surfaces and worldline formulations of imperfect boundaries extend the geometry and response~\cite{FoscoDirichlet,FoscoWorldline}. Reflectivity need not simply suppress production: a static, partially reflecting mirror with time-dependent properties can produce more particles than its perfectly reflecting counterpart in that model~\cite{RegoReflectivity}. Detectors select outgoing modes, while coherently driven plasma mirrors also involve incident radiation. These settings require the corresponding mode selection and boundary response; the present scalar totals alone do not determine a material-mirror yield.

Of course, particle number does not determine the entropy or correlations of the radiation~\cite{Holzhey1994,Bianchi2014,ChenYeom}. Interesting holographic mirrors and free-fermion models provide Page-curve and bulk-entropy constructions~\cite{AkalPage,ReyesPage} and partner-mode analyses distinguish Hawking-like correlations from a final energetic burst and examine purification through vacuum correlations~\cite{OsawaPartners,AgulloPurification}. Such questions require information about the outgoing state beyond $N$ and $E$.

Understanding how motion shapes radiation is useful beyond
the examples solved here. For example, carefully chosen cavity motion can avoid final particle creation~\cite{DelGrossoShortcut},
and exact solutions provide known answers against which
numerical methods can be tested~\cite{HsiungChen}. Our
solutions offer such reference cases with full relativistic
dependence, allowing particle number and energy to be
compared directly. We plan to extend this study in a separate
paper focusing on relativistic finite-location round trips.

\section{Conclusions}
\label{sec:conclusions}

For a real massless scalar field and a single perfectly reflecting mirror in $1+1$ dimensions, our main results are:
\begin{itemize}

\item To our knowledge, these are the first exact closed-form evaluations of finite, nonzero total particle number for smooth, asymptotically resting motions with full relativistic dependence. Moreover, the number and energy totals are elementary.

\item Cubic flight and symmetric bounce are unified within an exactly solvable two-parameter family, allowing unequal positive and negative velocity extrema and direct comparisons of their radiation.

\item For every fixed $m\ge1$ and time scale $\tau$, the polynomial flights obey
$N_m\sim C_m\eta_*\sim C_m\log\gamma$, while
$\tau E_m\sim D_m\gamma^{2+1/m}$.
Peak shape changes both particle multiplicity and the growth of the mean quantum energy.

\item Under sufficient uniform local and global hypotheses, the separated-event theorem extends these laws to general trajectories and establishes leading additivity. 

\item An exactly solvable elliptic count round trip has ultra-relativistic quadratic scaling $N\sim \eta^2$. Its different scaling illustrates a solution outside the theorem's hypotheses.

\end{itemize}
Together, these exact solutions and scaling laws show how the shape of relativistic mirror motion controls both the number of particles created from the vacuum and the energy they carry.

\begin{acknowledgments}
M.G. acknowledges partial support from Nazarbayev University in Qazaqstan through the FY2024-SGP-1-STMM Faculty Development Competitive Research Grant (FDCRGP), Nos.~201223FD8824 and SSH20224004. The work of E.I. is supported in part by the University Development
Fund of the Chinese University of Hong Kong, Shenzhen,
under Grant No. UDF01003869, UDF01003912, NSFC grant No. 12375080, and in part by the Ministry of Science and Technology of China under Grant No. 2024YFA1611004, and in part by the Natural
Science Foundation of Guangdong Province under Grant No. 2026A1515011242.

\section*{AI Statement}
The authors used OpenAI and Anthropic tools to assist with technical calculations. All the results were then checked by humans, and the presentation was completely rewritten by hand. The authors take full responsibility for the final manuscript. M.G. acknowledges Anthropic's access approval under its Claude Team plan for Scientists.

\end{acknowledgments}

\appendix

\section{Total particle and total energy representations}

In this Appendix we will derive the key formula \eqref{eq:pn-positive-number} used in the main text.
We also give a proof of the statement that the total energy, computed through the frequency representation \eqref{eq:observables} or through the energy representation \eqref{eq:stress} is the same. This statement is not new and has been proven before (as we shall discuss), but we believe the proof presented here may have a clearer structure and make the underlying assumptions more transparent.

\subsection{Particle number representation}
\label{app:positive}

Write $\Omega=\omega+\omega'$ and
$\mu=(\omega-\omega')/\Omega$, as in Eq.~\eqref{eq:frequency}.
Their ranges are $\Omega>0$ and $-1<\mu<1$, and their Jacobian is
$\dd\omega\dd\omega'=\Omega\dd\Omega\dd\mu/2$.
Equation~\eqref{eq:beta} becomes
\begin{equation}
 \begin{gathered}
 \beta^R_{\omega\omega'}=
 \frac{\sqrt{1-\mu^2}}{2\pi}J(\Omega),\\
 J(\Omega)=\int_\R v(t)e^{-i\Omega[t-\mu z(t)]}\dd t.
 \end{gathered}
 \label{eq:pn-beta-current}
\end{equation}
The relation $|\beta^R_{\omega\omega'}| = |\beta^L_{\omega'\omega}|$ gives $N=2N_R$. Consequently,
\begin{equation}
 N=\frac{1}{4\pi^2}\int_{-1}^1(1-\mu^2)\dd\mu
 \int_0^\infty\Omega|J(\Omega)|^2\dd\Omega.
 \label{eq:currentnorm}
\end{equation}
The factor of two here counts the two exterior regions; it is unrelated
to the restriction to positive Fourier frequencies below.

Below we will use the following functions:
\begin{equation}
 y=t-\mu z(t),\qquad
 \dd t = \frac{\dd y}{1 - \mu v(t)} , \qquad
 f(y)=\frac{v(t)}{1-\mu v(t)}.
 \label{eq:pn-ray-current}
\end{equation}
Note that $1-|\mu|\le\dd y/\dd t\le1+|\mu|$.
We use the Fourier transform convention
\begin{equation}
    \widehat f(\Omega)=\int_\R f(y)e^{-i\Omega y}\dd y.
\label{fourier_convension}
\end{equation}
Changing the integration variable in Eq.~\eqref{eq:pn-beta-current} from $t$ to $y$ according to Eq.~\eqref{eq:pn-ray-current}, we arrive at 
\begin{equation}
    J(\Omega) = \widehat f(\Omega).
\label{J_is_fhat}
\end{equation}
For slowly decaying currents, such as that of the symmetric bounce, we first apply Parseval's identity with a smooth cutoff in $y$ and then remove the cutoff in the resulting integrals.

Next, note that if Fourier transform of $f(y)$ is $\widehat f(\Omega)$, then Fourier transform of $f(y+h) - f(y)$ is $(e^{i \Omega h} - 1) \widehat f(\Omega)$. Applying Parseval's identity to this relation, we obtain:
\[
\begin{aligned}
&\int_{\mathbb R}|f(y+h)-f(y)|^2\,dy\\
&\quad=\frac1{2\pi}\int_{\mathbb R}
|e^{i\Omega h}-1|^2|\widehat f(\Omega)|^2\,d\Omega\\
&\quad=\frac1{2\pi}\int_{\mathbb R}
2[1-\cos(\Omega h)]|\widehat f(\Omega)|^2\,d\Omega.
\end{aligned}
\]
Now divide by \(h^2\) and integrate over \(h\). On the left, setting \(y'=y+h\) produces a functional which we call $G[f]$ (it will be important later):
\begin{equation}
 G[f]=\iint_{\R^2}\frac{[f(y)-f(y')]^2}{(y-y')^2}\dd y\dd y'.
 \label{eq:G}
\end{equation}
On the right, use
\[
\int_{\mathbb R}\frac{1-\cos(\Omega h)}{h^2}\,dh
=\pi|\Omega|
\]
and the fact that $f$ is real; this way, one arrives at
\begin{equation}
\iint_{\mathbb R^2}
\frac{|f(y)-f(y')|^2}{(y-y')^2}\,dy\,dy'
=
2\int_0^\infty\Omega|\widehat f(\Omega)|^2\,d\Omega
\label{f_key_identity}
\end{equation}
Let $\bar v(t,t')=[z(t)-z(t')]/(t-t')$. Direct subtraction gives
\begin{align}
 f(y)-f(y')
 &=\frac{v(t)-v(t')}{[1-\mu v(t)][1-\mu v(t')]},\nonumber\\
 y-y'&=(t-t')[1-\mu\bar v(t,t')].
 \label{eq:pn-difference-transform}
\end{align}

Plugging \eqref{eq:pn-difference-transform} and \eqref{J_is_fhat} into \eqref{f_key_identity}, changing the integration variables from $y,y'$ to $t,t'$ according to \eqref{eq:pn-ray-current}, and using the resulting identity, we can finally rewrite Eq.~\eqref{eq:currentnorm} as
\begin{equation}
 \begin{aligned}
 N={}&\frac{1}{8\pi^2}\int_{-1}^{1}(1-\mu^2)\,\mathrm{d}\mu\\
 &\times\iint_{\mathbb R^2}
 \frac{\left[\dfrac{v(t)-v(t')}{t-t'}\right]^2
 \,\mathrm{d}t\,\mathrm{d}t'}
 {[1-\mu v(t)][1-\mu v(t')][1-\mu\bar v(t,t')]^2}.
 \end{aligned}
 \label{eq:pn-positive-number_appendix}
\end{equation}
This is precisely the formula \eqref{eq:pn-positive-number} used in the main text.

We now discuss similar-in-spirit formulas that have appeared in the literature before, but to the best of our knowledge, the particle number representation \eqref{eq:pn-positive-number_appendix} in the context of the moving mirror model is written here explicitly for the first time. Below, we give credit to some related results found in the literature.

Notably, it was Ritus who transformed the spectral Bogoliubov norm into
double-null-coordinate integrals with a straight-line subtraction and a
principal-value prescription~\cite{Ritus1996Functional}, and subsequently
developed a proper-time representation~\cite{Ritus1999Vacuum}. These are
exact relativistic number functionals, although their integrands do not
display positivity as a squared difference. Fabbri, Navarro-Salas, and
Olmo obtained mode-resolved particle numbers from vacuum-subtracted
two-point functions~\cite{FabbriNavarroSalasOlmo2004}. Their expressions
involve derivatives of the ray map and outgoing-mode weights, whereas
the representation, Eq.~\eqref{eq:pn-positive-number_appendix}, directly expresses the total number as
an integral of a nonnegative kernel.

A related construction appears in classical radiation theory and is particularly
relevant in light of recent work on electron--mirror duality~\cite{IevlevGood2024ThermalLarmor,IevlevGoodLinder2024IR,
IevlevGood2023FermiDirac,IevlevGoodDavies2024Duality}. Stodolsky
expressed the mean number of emitted photons as a Lorentz-invariant
double-worldline integral and derived transverse-tangent representations
in which the coincident-point singularity is
removed~\cite{Stodolsky2002Photons}. Subsequent work examined inversion,
geometric properties, and the Euclidean counterpart of this
functional~\cite{Stodolsky2003Inversion,Stodolsky2006Curves,
RuijsenaarsStodolsky2008}. For rectilinear motion, classical photon
emission and mirror particle production are related by the established
mirror--charge correspondence~\cite{Ritus1996Functional,Ritus1999Vacuum}.
These results provide antecedents for an exact worldline description of
the integrated count. 
Equation~\eqref{eq:pn-positive-number_appendix} keeps the integral over $\mu$ explicit (retaining the frequency-asymmetry integration) and expresses the count in terms of a positivity-explicit integrand involving the squared velocity difference $[v(t)-v(t')]^2$ and the mirror's average velocity $\bar v(t,t')=[z(t)-z(t')]/(t-t')$.

\subsection{Energy from the particle spectrum}
\label{app:stress}

The renormalized stress tensor and its integrated positive-energy
form were obtained by Fulling and Davies~\cite{FD1976}, in particular
their Eqs.~(3.11) and~(4.5). We use that result rather than repeat its
derivation. Equality with the energy computed from Bogoliubov
coefficients is also already established in the literature: Walker rederived the radiated-energy
formula from the Bogoliubov transformation and examined the trajectory
condition needed for the relation to particle
production~\cite{WalkerEnergy}. Ritus gave a direct transformation of
the spectral energy into the stress-tensor expression, keeping track
of the endpoint, or Schott, terms~\cite[Sec.~3]{Ritus1996Functional}.
Related frequency transformations and Parseval arguments occur in
the mirror--charge correspondence~\cite{QuantumPower}.
We give a self-contained derivation of this established equality in
the variables of Appendix~\ref{app:positive}.

For an asymptotically resting smooth trajectory,
the known integrated
stress result, when its boundary terms vanish, is
\begin{align}
 E_R&=\frac{1}{12\pi}\int_\R
       \frac{\dot v^{\,2}}{(1-v)^3(1+v)^2}\dd t,\nonumber\\
 E_L&=\frac{1}{12\pi}\int_\R
       \frac{\dot v^{\,2}}{(1+v)^3(1-v)^2}\dd t,
 \label{eq:side-energy-functionals}
\end{align}
and hence
\begin{equation}
 E=E_R+E_L=\frac{1}{6\pi}\int_\R
                   \frac{\dot v^{\,2}}{(1-v^2)^3}\dd t.
 \label{eq:energy-positive-functional}
\end{equation}
These are the two reflected versions of the Fulling--Davies result
\cite{FD1976}; the two-sided proper-acceleration form is also given in
Ref.~\cite{QuantumPower}. The boundary terms vanish if
\begin{equation}
 \lim_{t\to\pm\infty}
 \frac{\dot v(t)}{[1\mp v(t)]^2[1\pm v(t)]}=0
 \quad\hbox{for both sign choices}.
 \label{eq:stress-boundaries}
\end{equation}
In particular, $v(t)\to0$ and $\dot v(t)\to0$ at both ends suffice.
We assume these limits when identifying the ordinary integrated
stress flux with Eq.~\eqref{eq:energy-positive-functional}. 

Now, let us present our own proof that the total energies computed from the particle spectrum in Eq.~\eqref{eq:observables} and from the representation Eq.~\eqref{eq:energy-positive-functional} are the same (assuming, of course, that all integrals converge and give finite and well-defined results).
First write the energy on the right as
\begin{equation}
 E_R=\int_0^\infty\!\int_0^\infty
       \omega|\beta^R_{\omega\omega'}|^2\dd\omega\dd\omega'.
 \label{eq:energy-right-spectrum}
\end{equation}
The relation
$|\beta^L_{\omega\omega'}|=|\beta^R_{\omega'\omega}|$
allows us to exchange the two frequency labels in the left energy.
It therefore gives
\begin{equation}
 E_L=\int_0^\infty\!\int_0^\infty
       \omega'|\beta^R_{\omega\omega'}|^2\dd\omega\dd\omega'.
 \label{eq:energy-left-spectrum}
\end{equation}
The weight in the sum is consequently $\omega+\omega'=\Omega$,
even when the two side energies differ.

As in Appendix~\ref{app:positive}, use
\begin{equation}
 \begin{gathered}
 \omega=\frac{\Omega(1+\mu)}2,\qquad
 \omega'=\frac{\Omega(1-\mu)}2,\\
 \dd\omega\dd\omega'=\frac{\Omega}{2}\dd\Omega\dd\mu.
 \end{gathered}
 \label{eq:energy-frequency-change}
\end{equation}
Equation~\eqref{eq:pn-beta-current} supplies
$|\beta^R|^2=(1-\mu^2)|J(\Omega)|^2/(4\pi^2)$.
Multiplying this by the energy weight and the Jacobian gives
\begin{equation}
 E=\frac{1}{8\pi^2}\int_{-1}^1(1-\mu^2)\dd\mu
                   \int_0^\infty\Omega^2|J(\Omega)|^2\dd\Omega.
 \label{eq:energy-current-spectrum}
\end{equation}

We use the same definitions of $y$ and $f(y)$ as in Eq.~\eqref{eq:pn-ray-current} and the relation $J=\widehat f$, see Eq.~\eqref{J_is_fhat}.

The derivative of $f$ is given by
\begin{align}
 \frac{\partial f}{\partial y}
 &=\frac{1}{1-\mu v}\frac{\dd}{\dd t}
                     \left(\frac{v}{1-\mu v}\right)\nonumber\\
 &=\frac{\dot v}{(1-\mu v)^3}.
 \label{eq:energy-current-derivative}
\end{align}
With the Fourier convention \eqref{fourier_convension}, the
transform of $\partial_yf$ is $i\Omega\widehat f$.
Parseval's identity and the reality of $f$ now give
\begin{align}
 \int_0^\infty\Omega^2|\widehat f(\Omega)|^2\dd\Omega
 &=\pi\int_\R|\partial_yf(y)|^2\dd y\nonumber\\
 &=\pi\int_\R\frac{\dot v^{\,2}}{(1-\mu v)^5}\dd t.
 \label{eq:energy-parseval}
\end{align}
The last equality uses $\dd y=(1-\mu v)\dd t$.

Inserting Eq.~\eqref{eq:energy-parseval} into
Eq.~\eqref{eq:energy-current-spectrum} gives
\begin{equation}
 E=\frac{1}{8\pi}\int_{-1}^1(1-\mu^2)\dd\mu
                  \int_\R\frac{\dot v^{\,2}}{(1-\mu v)^5}\dd t.
 \label{eq:spectral-energy}
\end{equation}
The integral over $\mu$ is elementary,
\begin{equation}
 \int_{-1}^1\frac{1-\mu^2}{(1-\mu v)^5}\dd\mu
       =\frac{4}{3(1-v^2)^3}.
 \label{eq:energy-angular}
\end{equation}
Equations~\eqref{eq:spectral-energy}
and~\eqref{eq:energy-angular} yield
Eq.~\eqref{eq:energy-positive-functional}, including its normalization.
Under the stated endpoint conditions, this spectral energy is
therefore exactly the total integrated stress energy in
Eq.~\eqref{eq:stress}.

\section{Exact calculations for flights and returns}
\label{app:exact}

In this Appendix we compute particle counts and total energies for various trajectories covered in the main text.

\subsection{Generalized Airy moment}
\label{app:moment}

We evaluate $J_{2m+1}$, Eq.~\eqref{eq:airyfamily} directly from that defining integral.

For each fixed positive integer $m$, our convention is
\begin{equation}
 \mathcal A_{2m+1}(x)=\frac1\pi\int_0^\infty
 \cos\left(\frac{s^{2m+1}}{2m+1}+xs\right)\dd s,
\end{equation}
which is equivalent to Eq.~\eqref{eq:airyfamily}. An equivalent contour
representation is given in
Ref.~\cite[Eq.~(2.5)]{BeteaBouttierWalsh2021}.
Differentiation and an integration by parts in $s$ give
\begin{equation}
 \mathcal A_{2m+1}^{(2m)}(x)
 =(-1)^{m+1}x\mathcal A_{2m+1}(x).
 \label{eq:ode}
\end{equation}
This is the generalized Airy differential equation stated immediately
after Eq.~(2.5) of Ref.~\cite{BeteaBouttierWalsh2021}.

To obtain the moment, introduce the auxiliary function
$g(x)=x\mathcal A_{2m+1}'(x)$. Leibniz's rule and
Eq.~\eqref{eq:ode} give
\begin{equation}
 \begin{aligned}
 g^{(2m)}(x)={}&(-1)^{m+1}\bigl[x^2\mathcal A_{2m+1}'(x)\\
 &\qquad +(2m+1)x\mathcal A_{2m+1}(x)\bigr].
 \end{aligned}
\end{equation}
We now use the Lagrange integration-by-parts identity underlying
the higher-order Airy Christoffel--Darboux calculation in
Ref.~\cite[Supplemental Material, Sec.~D,
Eqs.~(52)--(54)]{LeDoussalMajumdarSchehr2018}.
In the present case its form is
\begin{align}
 &\frac{\dd}{\dd x}\sum_{\ell=0}^{2m-1}(-1)^\ell
 \mathcal A_{2m+1}^{(\ell)}(x)
 g^{(2m-1-\ell)}(x)\nonumber\\
 &\quad=\mathcal A_{2m+1}(x)g^{(2m)}(x)
       -\mathcal A_{2m+1}^{(2m)}(x)g(x)\nonumber\\
 &\quad=(-1)^{m+1}(2m+1)x\mathcal A_{2m+1}(x)^2.
\label{long_airy_identity}
\end{align}
The first equality follows by cancellation of adjacent terms;
the second uses Eq.~\eqref{eq:ode}.

Now integrate \eqref{long_airy_identity} from zero to infinity. At zero,
$g^{(j)}(0)=j\mathcal A_{2m+1}^{(j)}(0)$ for $j\ge0$,
including $g(0)=0$. Pairing the summands with indices $\ell$
and $2m-1-\ell$ therefore yields for the moment $J_{2m+1}$ introduced in Eq.~\eqref{eq:airyfamily}:
\begin{align}
 J_{2m+1}
 ={}&\frac{(-1)^m}{2m+1}\sum_{\ell=0}^{m-1}
 (-1)^\ell(2m-1-2\ell)\nonumber\\
 &\times\mathcal A_{2m+1}^{(\ell)}(0)
          \mathcal A_{2m+1}^{(2m-1-\ell)}(0).
 \label{eq:momentendpoint}
\end{align}
The derivatives at zero follow from the defining integral Eq.~\eqref{eq:airyfamily}; in this case, that integral reduces to the Gamma function. We obtain
\begin{align}
 \mathcal A_{2m+1}^{(\ell)}(0)
 ={}&\frac{(2m+1)^{(\ell+1)/(2m+1)-1}}{\pi}
 \Gamma\left(\frac{\ell+1}{2m+1}\right)\nonumber\\
 &\times\cos\left[\frac{\pi\ell}{2}
       +\frac{\pi(\ell+1)}{2(2m+1)}\right].
 \label{eq:jets}
\end{align}

To simplify the products of the Gamma function which now occur in Eq.~\eqref{eq:momentendpoint}, 
note that the two gamma arguments in Eq.~\eqref{eq:momentendpoint} sum to
one, and the two cosine arguments sum to $m\pi$.
Using Euler's reflection formula
\cite[Eq.~5.5.3]{DLMF},
\begin{equation}
 \Gamma(a)\Gamma(1-a)=\frac{\pi}{\sin(\pi a)},
\end{equation}
together with
\[
 (-1)^\ell\cos^2\left(\frac{\pi\ell}{2}+\frac\theta2\right)
 =\frac{\cos\theta+(-1)^\ell}{2},
\]
we obtain
\begin{equation}
 \begin{aligned}
 &(-1)^\ell\mathcal A_{2m+1}^{(\ell)}(0)
                 \mathcal A_{2m+1}^{(2m-1-\ell)}(0)\\
 &\qquad=\frac{(-1)^m}{2\pi(2m+1)}
             \left[\cot\theta + \frac{(-1)^\ell}{\sin \theta} \right].
 \end{aligned}
\end{equation}
where $\theta=\pi(\ell+1)/(2m+1)$.
Substituting the derivative products in
Eq.~\eqref{eq:momentendpoint} and writing $p=\ell+1$ gives
\begin{align}
 J_{2m+1}={}&\frac1{2\pi(2m+1)^2}
 \sum_{p=1}^{m}(2m+1-2p)\nonumber\\
 &\times\left[\cot\frac{\pi p}{2m+1}
       +\frac{(-1)^{p-1}}{\sin[\pi p/(2m+1)]}\right].
 \label{eq:moment-weighted-sum}
\end{align}
We now simplify this intermediate sum without changing the endpoint
calculation. Put $\theta_p=\pi p/(2m+1)$ for $1\le p\le m$.
For even $p=2j$, the half-angle identity gives
\begin{equation}
 \cot\theta_p-\frac1{\sin\theta_p}
   =2\cot\theta_p-\cot\theta_j.
 \label{eq:moment-even-reindexing}
\end{equation}
For odd $p=2m+1-2j$, one has
$\theta_j=\pi/2-\theta_p/2$, and hence
\begin{equation}
 \cot\theta_p+\frac1{\sin\theta_p}
   =2\cot\theta_p+\cot\theta_j.
 \label{eq:moment-odd-reindexing}
\end{equation}
The even indices give $j=1,\ldots,\lfloor m/2\rfloor$;
the odd indices give the remaining integers
$j=\lfloor m/2\rfloor+1,\ldots,m$ in reverse order. Together
they therefore give each $j\in\{1,\ldots,m\}$ exactly once.
In both cases the signed weight multiplying $\cot\theta_j$ is
$4j-(2m+1)$: it is $-(2m+1-4j)$ in the even case and
$(2m+1)-2(2m+1-2j)$ in the odd case.
Consequently,
\begin{align}
 &\sum_{p=1}^{m}(2m+1-2p)
    \left[\cot\theta_p+\frac{(-1)^{p-1}}{\sin\theta_p}\right]
       \nonumber\\
 &\quad=2\sum_{p=1}^{m}(2m+1-2p)\cot\theta_p
       \nonumber\\
 &\qquad+\sum_{j=1}^{m}[4j-(2m+1)]\cot\theta_j\nonumber\\
 &\quad=(2m+1)\sum_{p=1}^{m}\cot\theta_p.
 \label{eq:moment-sum-reduction}
\end{align}
Equation~\eqref{eq:moment-weighted-sum} thus becomes
\begin{equation}
 J_{2m+1}=\frac1{2\pi(2m+1)}
            \sum_{p=1}^{m}\cot\frac{\pi p}{2m+1}.
 \label{eq:Jd}
\end{equation}
Finally, multiplication by $2(2m+1)/m$ gives directly
\begin{equation}
 C_m=\frac1{\pi m}\sum_{p=1}^{m}
                      \cot\frac{\pi p}{2m+1},
\end{equation}
which is Eq.~\eqref{eq:Cm}. Each angle lies strictly between
zero and $\pi/2$, so every summand is positive.

For $m=1$, this reduces to
\begin{equation}
 J_3=-\frac{\operatorname{Ai}(0)\operatorname{Ai}'(0)}3
     =\frac1{6\pi\sqrt3}.
\end{equation}
The same result follows directly from the standard Airy-product
integral \cite[Eq.~9.11.8]{DLMF},
\begin{equation}
 \begin{aligned}
 \int x\operatorname{Ai}(x)^2\dd x
 ={}&\frac13\bigl[x^2\operatorname{Ai}(x)^2-x\operatorname{Ai}'(x)^2\\
 &\qquad+\operatorname{Ai}(x)\operatorname{Ai}'(x)\bigr].
 \end{aligned}
\end{equation}

\subsection{Polynomial-flight number and energy}
\label{app:flightenergy}

Introduce $y=-z/(V\tau)$. Along the polynomial flight,
\begin{equation}
 \begin{gathered}
 t=\tau\left(y+\frac{y^{2m+1}}{2m+1}\right),\\
 v=-\frac{V}{1+y^{2m}},\qquad
 \dot v=\frac{2mV y^{2m-1}}{\tau(1+y^{2m})^3}.
 \end{gathered}
 \label{eq:flight-variables}
\end{equation}
The first formula is strictly increasing in $y$. Substitution in
Eq.~\eqref{eq:beta}, followed by
$y=(\tau\Omega)^{-1/(2m+1)}s$, gives exactly
Eq.~\eqref{eq:familybeta}, including its overall normalization.
In the squared coefficient set
$x=(1+V\mu)(\tau\Omega)^{2m/(2m+1)}$.
The resulting frequency integral is
\begin{equation}
 N_m=\frac{2m+1}{2m}J_{2m+1}V^2
       \int_{-1}^{1}\frac{1-\mu^2}{(1+V\mu)^2}\dd\mu.
 \label{eq:flight-number-integral}
\end{equation}
The angular factor is a textbook rational function integral:
\begin{align}
 V^2\int_{-1}^1\frac{1-\mu^2}{(1+V\mu)^2}\dd\mu
 =4\left(\frac{\atanh V}{V}-1\right)=4F(\eta_*).
 \label{eq:flight-angular}
\end{align}
This proves Eq.~\eqref{eq:familyN}, and includes the cubic count.

For the total energy, the stress tensor result \eqref{eq:stress} gives
\begin{equation}
 E_m=\frac{4m^2V^2}{6\pi\tau}\int_\R
 \frac{y^{4m-2}(1+y^{2m})}
 {[(1+y^{2m})^2-V^2]^3}\dd y.
 \label{eq:flightintegral}
\end{equation}
Changing the integration variable to $h=y^{2m}$, we obtain
\begin{equation}
 E_m=\frac{2mV^2}{3\pi\tau}\int_0^\infty
 \frac{h^{1-1/(2m)}(1+h)}{[(1+h)^2-V^2]^3}\dd h.
\end{equation}
Now
$\partial_h\{[(1+h)^2-V^2]^{-2}\}
=-4(1+h)[(1+h)^2-V^2]^{-3}$.
The integration-by-parts boundary term is proportional to
$h^{1-1/(2m)}/[(1+h)^2-V^2]^2$ and vanishes at both ends.
Consequently,
\begin{equation}
 \begin{aligned}
 E_m={}&\frac{2mV^2[1-1/(2m)]}{12\pi\tau}\\
 &\times
 \int_0^\infty
 \frac{h^{-1/(2m)}}{(h+1-V)^2(h+1+V)^2}\dd h.
 \end{aligned}
 \label{eq:energybeta}
\end{equation}
For $V>0$, we can rewrite it in terms of the partial fractions:
\begin{equation}
 \begin{gathered}
 \frac1{(h+1-V)^2(h+1+V)^2}
 \\
 \begin{aligned}
 ={}&\frac{(h+1-V)^{-2}+(h+1+V)^{-2}}{4V^2}\\
 &+\frac{(h+1+V)^{-1}-(h+1-V)^{-1}}{4V^3}.
 \end{aligned}
 \end{gathered}
 \label{eq:fractions}
\end{equation}
For $c>0$, we use the following beta-function integrals:
\begin{align}
 \int_0^\infty\frac{h^{-1/(2m)}}{h+c}\dd h
 &=\frac{\pi c^{-1/(2m)}}{\sin[\pi/(2m)]},\nonumber\\
 \int_0^\infty\frac{h^{-1/(2m)}}{(h+c)^2}\dd h
 &=\frac{\pi c^{-1-1/(2m)}}{2m\sin[\pi/(2m)]}.
 \label{eq:flight-beta-integrals}
\end{align}
Inserting $c=1-V,1+V$ proves Eq.~\eqref{eq:familyE}.
At $m=1$, combining the square roots with
$\sqrt{1-V}\sqrt{1+V}=\sqrt{1-V^2}$ gives the manifestly
regular cubic expression Eq.~\eqref{eq:cubicE}.
The apparent singularity at $V=0$ in the general formula is only a
partial-fraction cancellation. Directly from
Eq.~\eqref{eq:flightintegral},
\begin{equation}
 E_m=\frac{2m}{3\pi\tau}
 B\left(2-\frac1{2m},3+\frac1{2m}\right)V^2+O(V^4).
 \label{eq:flight-low-speed}
\end{equation}
For $m=1$ this is $5V^2/(192\tau)+O(V^4)$.

The count has the expansion
$F(\eta_*)=\eta_*-1+o(1)=\log\gamma+\log2-1+o(1)$
as $V\to1$. The sole leading singular term of the energy is
$(1-V)^{-1-1/(2m)}$, with the coefficient already evaluated in
the companion theorem proof~\cite{MirrorCountsGitHub}. These observations prove both forms of
Eq.~\eqref{eq:familylimit}.

\subsection{Two-parameter cubic motion}
\label{app:clock}

For the trajectory Eq.~\eqref{eq:clockexplicit}, use the dimensionless
integration coordinate $y$ defined by
\begin{equation}
 t=\tau(y+y^3/3),\qquad
 z=\tau[(V_+-V_-)y+\sqrt{V_+V_-}\,y^2].
 \label{eq:clock-parameter}
\end{equation}
The first relation is globally invertible, and its inverse is
$y=2\sinh[\operatorname{arsinh}(3t/(2\tau))/3]$.

\subsubsection{Airy amplitude and exact number}

To compute the Bogoliubov beta coefficients, it is convenient to shift the trajectory parameter $y$ as $y=\chi+\mu\sqrt{V_+V_-}$. 
We start with the formula for the beta coefficients in Eq.~\eqref{eq:beta}.
In terms of $\chi$, the phase of the exponential becomes
\begin{align}
 \frac{t-\mu z}{\tau}&=\frac{\chi^3}{3}+D(\mu)\chi+\Phi(\mu),\nonumber\\
 D(\mu)&=(1-V_+\mu)(1+V_-\mu),\nonumber\\
 \Phi(\mu)&=\mu\sqrt{V_+V_-}[1-\mu(V_+-V_-)]
              -\tfrac23\mu^3(V_+V_-)^{3/2}.
 \label{eq:clockD}
\end{align}
Now, the integration measure becomes
\begin{equation}
    v \dd t = \dd z=\tau[\mathcal B(\mu)+2\sqrt{V_+V_-}\chi]\dd\chi
\end{equation}
where $\mathcal B=V_+-V_-+2V_+V_-\mu=-D'$.
The beta Bogoliubov integral splits in two:
\begin{align}
 \int_\R e^{-i\tau\Omega(\chi^3/3+D\chi)}\dd\chi
 &=2\pi (\tau\Omega)^{-1/3}\Ai(D(\tau\Omega)^{2/3}),\nonumber\\
 \int_\R\chi e^{-i\tau\Omega(\chi^3/3+D\chi)}\dd\chi
 &=2\pi i (\tau\Omega)^{-2/3}\Ai'(D(\tau\Omega)^{2/3}).
\end{align}
The second follows by differentiating the first with respect to $D$.
Thus
\begin{align}
 \beta^R={}&e^{-i\tau\Omega\Phi}\tau\sqrt{1-\mu^2}
 \bigl[\mathcal B (\tau\Omega)^{-1/3}\Ai(D(\tau\Omega)^{2/3})\nonumber\\
 &\hspace{10mm}+2i\sqrt{V_+V_-}\,(\tau\Omega)^{-2/3}\Ai'(D(\tau\Omega)^{2/3})\bigr].
 \label{eq:clockbeta}
\end{align}

Now, the particle count integral involves $|\beta^R|^2$. When we compute this absolute square, there is no cross term from \eqref{eq:clockbeta}.
The Airy equation gives
\begin{equation}
 \int_0^\infty(\Ai')^2\dd x
 =-\Ai(0)\Ai'(0)-\int_0^\infty x\Ai^2\dd x=2J_3.
\end{equation}
Then, the particle count integral is reduced to
\begin{equation}
 N_{\rm c}=\frac{3J_3}{2}\int_{-1}^1(1-\mu^2)
 \left(\frac{\mathcal B^2}{D^2}+\frac{8V_+V_-}{D}\right)\dd\mu.
 \label{eq:clockintegral}
\end{equation}
This may look a bit complicated, but after plugging back the definitions above, it reduces to an ordinary rational function integral, and the result is precisely Eq.~\eqref{eq:clockN}, with the symmetric bounce \eqref{eq:bouncetotals} and cubic flight \eqref{eq:cubicN} as the stated special cases.

\subsubsection{Total energy}

Computation of the total radiated energy is more straightforward, because we have a simple stress-energy result \eqref{eq:stress}.
All we need to do is substitute the velocity.
This can be done by first differentiating  Eq.~\eqref{eq:clock-parameter}, which gives
\begin{equation}
 v(y)=\frac{V_+-V_-+2\sqrt{V_+V_-}\,y}{1+y^2}.
 \label{eq:clock-velocity}
\end{equation}
Plugging that into Eq.~\eqref{eq:stress} again gives a textbook rational integral. We will not bother the reader with a straightforward but tedious calculation, but simply present the result. In terms of the maximal rapidities \eqref{eq:clockpeaks},
\begin{equation}
\begin{aligned}
E_{\rm c}={}&
\frac{(\cosh\eta_+\cosh\eta_-)^{3/2}}{48\tau}
\cosh\!\left(\frac{\eta_+-\eta_-}{2}\right)
\\
&\times\Bigg[
4\cosh^2\!\left(\frac{\eta_+-\eta_-}{2}\right)-3
+\frac{5}{\cosh^2\!\left(\frac{\eta_++\eta_-}{2}\right)}
\\
&\qquad-
\frac{6\cosh^2\!\left(\frac{\eta_+-\eta_-}{2}\right)}
{\cosh^4\!\left(\frac{\eta_++\eta_-}{2}\right)}
\Bigg].
\end{aligned}
\end{equation}
In terms of the peak speeds themselves,
\begin{widetext}
\begin{equation}
E_{\rm c}=
\frac{
 \sqrt{1-V_+V_-+\sqrt{(1-V_+^2)(1-V_-^2)}}
}{
 48\sqrt{2}\,\tau
 \bigl[(1-V_+^2)(1-V_-^2)\bigr]^{3/2}
}
\left[
\begin{aligned}
&2(1-V_+V_-)-\sqrt{(1-V_+^2)(1-V_-^2)}
\\
&+\frac{
 2(1-V_+^2)(1-V_-^2)
 \left[11V_+V_--1-\sqrt{(1-V_+^2)(1-V_-^2)}\right]
}{
 \left[1+V_+V_-+\sqrt{(1-V_+^2)(1-V_-^2)}\right]^2
}
\end{aligned}
\right].
\label{eq:clockE-speeds}
\end{equation}
\end{widetext}

\section{An elliptic count round-trip mirror}
\label{app:pendulum}

It's important to note that not every mirror that comes to rest obeys the logarithmic ultra-relativistic law, $N \propto \log \gamma$.
Here, we record a round trip whose totals are exact and whose
count grows faster; $N \propto (\log \gamma)^2$.
In null coordinates $u=t-z$ and $v=t+z$, the mirror follows
\begin{equation}
  (v-u)(\tau^2+uv)=2L\,uv .
  \label{eq:pendtraj}
\end{equation}
Here $L$ is the max excursion length. The mirror actually
starts at rest at $L$, turns around at $z=0$, and returns
to $L$. Here $\tau$ is our usual time-scale parameter,
not the duration of the excursion. Consequently, $L/\tau$
need not be less than unity, although the physical speed
is always $|\dot z|<1$.

Including both sides, we find the energy radiated and
the particle count:
\begin{equation}
  E=\frac{L^2}{8\tau^3},\qquad
  N=\frac{4}{\pi^2}\,\mathsf{K}(k)
    \bigl[\mathsf{K}(k)-\mathsf{E}(k)\bigr],
  \label{eq:pendtotals}
\end{equation}
where $k=L/\sqrt{L^2+4\tau^2}$.
Here $\mathsf{K}$, $\mathsf{E}$ are the complete elliptic
integrals in the modulus convention. The energy is exactly
quadratic in $L$ at fixed $\tau$; there are no relativistic
corrections. With $u=\tau\tan(\phi/2)$ and
$v=\tau\tan(\psi/2)$ the trajectory is simply:
$\cot\phi-\cot\psi=L/\tau$.

In the ultra-relativistic regime, there is some interesting
physics. As $L/\tau\to\infty$, the peak Lorentz factor is
$\gamma\simeq L/\tau$, and
\begin{equation}
  N=\frac{4}{\pi^2}(\log\gamma)^2+O(\log\gamma),\qquad
  E\simeq\frac{\gamma^2}{8\tau},
  \label{eq:pendlimit}
\end{equation}
or $N=(4/\pi^2)\,\eta^2+O(\eta)$ in terms of the peak
rapidity $\eta$. The count grows as the square of the
logarithm rather than as $C_1\log\gamma$ for cubic flight
in Eq.~\eqref{eq:cubiclimit}. This does not contradict
the local scaling result in Sec.~\ref{sec:local}.
The mirror lingers near the speed of light for a time
of order $L$ while its turn sharpens, so the quadratic
coefficient $a$ defined below tends to zero and the
uniform assumptions of that result do not hold.

Explicitly, using units with $\tau=1$, the velocity
near its positive-time maximum is
\begin{equation}
  \dot z(t)=V-a(t-t_*)^2+O((t-t_*)^3),
\end{equation}
where $V=\dot z(t_*)$. As $L\to\infty$,
\begin{equation}
  t_*\sim(L/4)^{1/3},
  \qquad
  a\sim\frac{3}{2^{1/3}L^{10/3}}\to0.
\end{equation}
The coefficient $a$ is therefore not bounded away from
zero in this limit, so the local logarithmic scaling
result does not apply.

\bibliography{sources}
\end{document}